\documentclass[twocolumn,twoside]{revtex4}
\usepackage{graphicx}
\usepackage{fancyhdr}
\begin{document}

\title{Measurements of $\gamma$ from tree-level decays}

%

\author{A. Bertolin \\
on behalf of the LHCb Collaboration}
\affiliation{Istituto Nazionale di Fisica Nucleare, Sezione di Padova, Italy}
%


\begin{abstract}
The LHCb approach for a precise determination of the CKM angle $\gamma$ from tree--level 
decays is presented. Up to 16 independent determinations, using a variety of beauty and
charm mesons decay modes, are available at LHCb. Not all of them have the same sensitivity 
to $\gamma$. The best accuracy is reached by combining all of the available results.
These proceedings review some of the independent determinations used in the combination
and present the latest combined result available, as of Summer 2018.
\end{abstract}

\maketitle

\thispagestyle{fancy}


\section{Introduction}
Following the arguments developed in \cite{ref:LHCb-gamma-pap}, one of the mandatory 
requirements to understand the origin of the baryon asymmetry of the Universe is that 
both charge (C) and charge-parity (CP) symmetries are broken.
The latter phenomenon arises in the Standard Model (SM) of particle physics through the 
complex phase of the Cabibbo-Kobayashi-Maskawa (CKM) quark mixing matrix, although 
the effect in the SM is not large enough to account for the observed baryon asymmetry in 
the Universe.
Violation of CP symmetry can be studied by measuring the angles of the CKM unitarity triangle.
One of these angles,
$\gamma \equiv \arg( - V^{\phantom{\star}}_{ud} V^{\star}_{ub} / V^{\phantom{\star}}_{cd} V^{\star}_{cb}$),
can be measured using only tree--level processes; a method that, assuming new physics is not 
present in tree-level decays, has negligible theoretical uncertainty.
Disagreement between such direct measurements of $\gamma$ and the value inferred from global 
CKM fits, assuming the validity of the SM, would indicate new physics beyond the SM.
The value of $\gamma$ can be determined by exploiting the interference between favoured
$b \rightarrow c W$ ($V_{cb}$) and suppressed $b \rightarrow u W$ ($V_{ub}$) transition
amplitudes.
The most precise way to determine $\gamma$ is through a combination of measurements from 
many decay modes. Up to 16 decay modes are considered at present by the
LHCb Collaboration \cite{ref:LHCb-CONF-2018-002}.
Three of them, using very different analysis techniques, will be briefly discussed in
the next sections. The last but one section will present the results of the combination. 
The last will draw conclusions and future prospects.

\section{$B^+ \rightarrow D K^+ (D \rightarrow K^0_{\text{S}} h^+ h^-)$ analysis}

\begin{figure}[h]
\centering
\includegraphics[width=85mm]{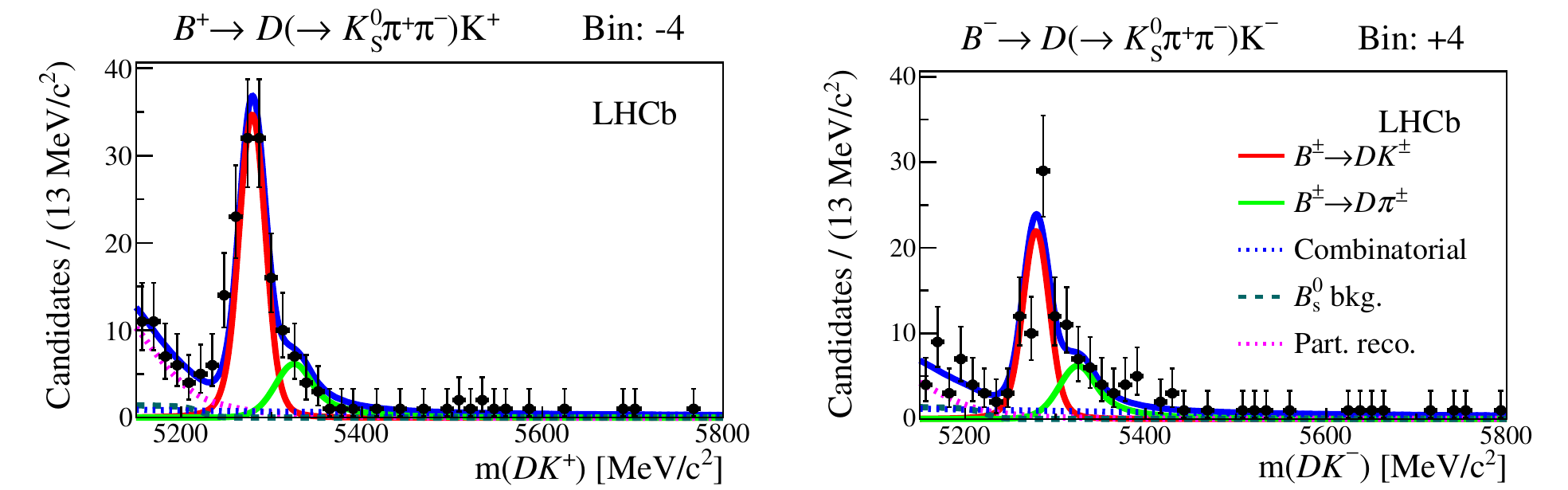}
\caption{Invariant mass distributions of (left) $B^+ \rightarrow D K^+$ and (right) 
$B^- \rightarrow D K^-$ in bin -4 and 4, respectively.
Reproduced from Ref. \cite{ref:input1}.}
\label{fig:input1_1}
\end{figure}

\begin{figure}[h]
\centering
\includegraphics[width=60mm]{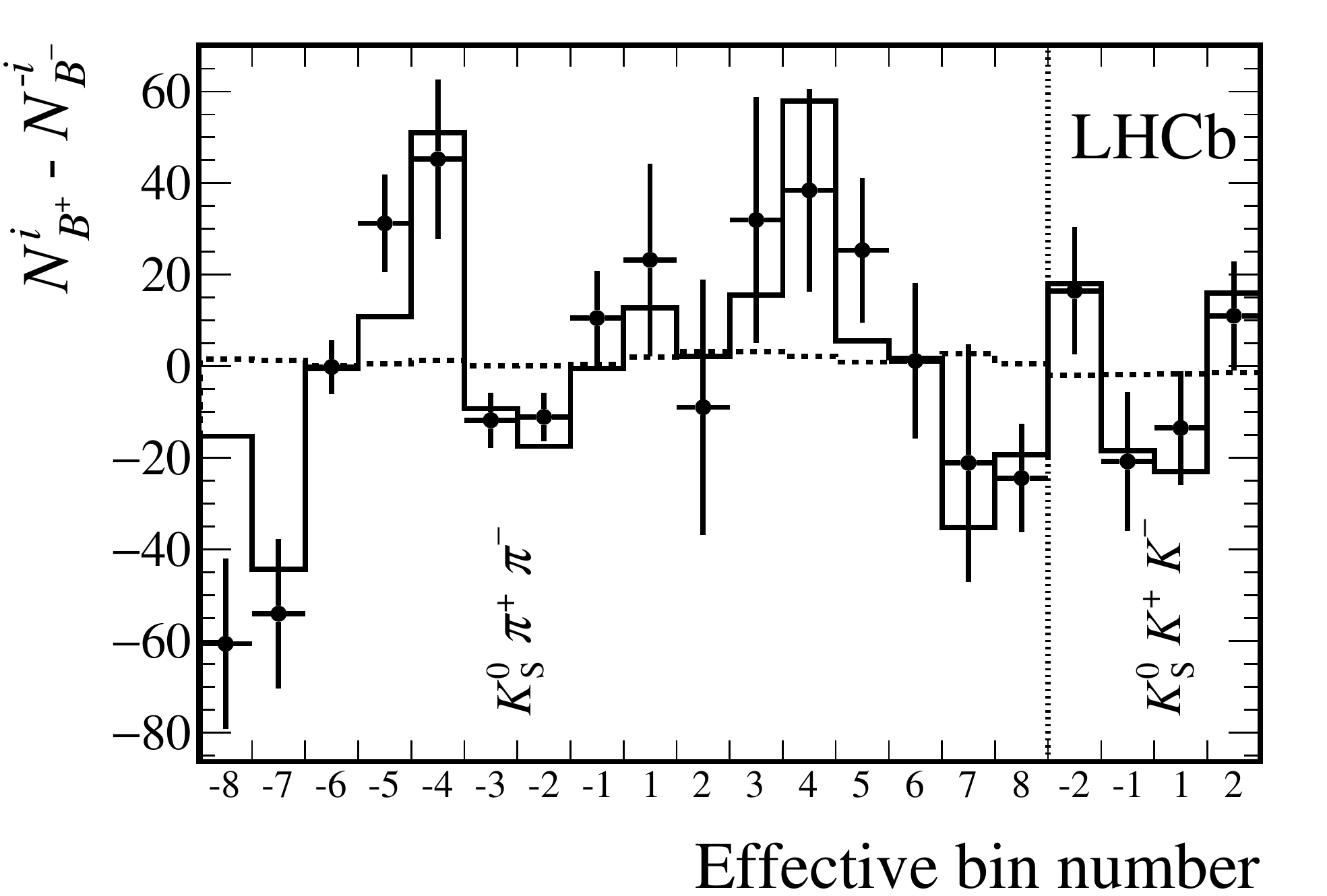}
\caption{Difference in the $B^+$ and $B^-$ yields as a function of the effective bin number.
Reproduced from Ref. \cite{ref:input1}.} 
\label{fig:input1_2}
\end{figure}

\begin{figure}[h]
\centering
\includegraphics[width=60mm]{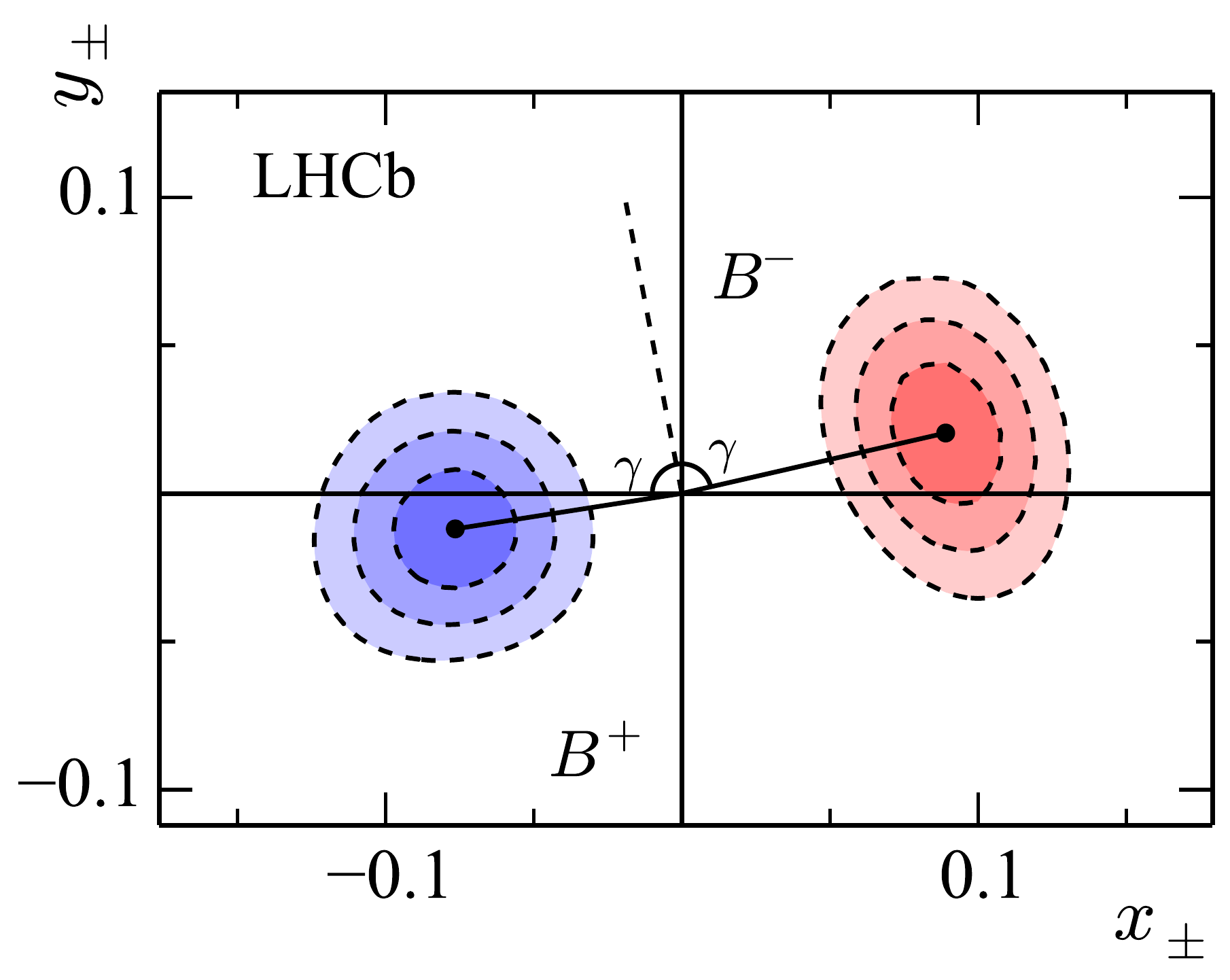}
\caption{Graphical representation of the $x_{\pm}$ and $y_{\pm}$ values in terms of 
$\gamma$. The 68.2 \%, 95.5 \% and 99.7 \% probability regions for $x_{\pm}$ and $y_{\pm}$
are represented by the dashed areas. Reproduced from Ref. \cite{ref:input1}.}
\label{fig:input1_3}
\end{figure}

In the decay mode $B^+ \rightarrow D K^+ (D \rightarrow K^0_{\text{S}} h^+ h^-)$, where $D$
represents a neutral charm meson that is a mixture of the $D^0$ and $\overline{D}^0$
flavour eigenstates and $h$ a $\pi$ or a $K$ meson, the sensitivity to $\gamma$ is obtained 
comparing the $D$ meson Dalitz plot distribution for reconstructed $B^+$ and 
$B^-$ decays \cite{ref:input1}.
The $B^- \rightarrow D K^-$ decay amplitude can be written as: \\
$A_B(m^2_-,m^2_+) \propto A_D(m^2_-,m^2_+) +$ \\
$r_B e^{i(\delta_B-\gamma)} A_{\overline{D}}(m^2_-,m^2_+)$ \\
a sum of a favoured, the first, and a suppressed, the latter, amplitudes, where $m^2_{\pm}$ 
represents the $K^0_{\text{S}} h^{\pm}$ invariant mass squared and $r_B$ ($\delta_B$) is the ratio (strong 
phase difference) between these amplitudes. Here and in the following the quantities
labeled just $r_B$ and $\delta_B$ are unique to each $B$ decay mode.
Four CPV observables defined as: \\
$x_{\pm} \equiv r_B \cos(\delta_B \pm \gamma)$ and $y_{\pm} \equiv r_B \sin(\delta_B \pm \gamma)$ \\
are determined by measuring the $B^+$ and $B^-$ yields in bins of the Dalitz plot variables 
$m^2_+$ and $m^2_-$.
The strong phase difference between the $D^0$ and $\overline{D}^0$ amplitudes at a given point 
of the Dalitz plot is also needed. This phase difference is being directly measured by the CLEO 
collaboration exploiting quantum--correlated $D^0$ $\overline{D}^0$ pairs produced at the $\psi(3770)$ 
resonance \cite{ref:CLEO}.
This approach makes the measurement independent of modelling the $D$ decay amplitude.
The $D$--decay Dalitz plot is partitioned into 2 $\times$ $N$ bins labelled from $i = -N$ to
$i = +N$ (excluding zero), symmetric around $m^2_- = m^2_+$ such that if $(m^2_-,m^2_+)$
is in bin $i$ then $(m^2_+,m^2_-)$ is in bin $-i$. By convention, the positive values of $i$
correspond to bins for which $m^2_- > m^2_+$.
The partition was chosen to optimise the statistical sensitivity to $\gamma$.
The $B^{\pm} \rightarrow D K^{\pm}$ invariant mass distributions obtained in bin -4 and
4, using $D \rightarrow K^0_{\text{S}} \pi^+ \pi^-$ and the 2015--2016 LHCb data are shown 
in Fig. \ref{fig:input1_1}: a clear asymmetry is visible. The data points are fitted using
a signal and several background components. 
Overall a yield of about 2000 events is observed for each of $B^- \rightarrow D K^-$ and 
$B^+ \rightarrow D K^+$.
The difference in the $B^+$ and $B^-$ yields as a function of this effective bin number
is shown in Fig. \ref{fig:input1_2}. Dots represent data, the dotted line the expectation
without CPV and the continous line the fit prediction with the central values of the
parameters $x_{\pm}$ and $y_{\pm}$. 
A graphical representation of the $x_{\pm}$ and $y_{\pm}$ values in terms of $\gamma$
is given in Fig. \ref{fig:input1_3}, corresponding to:
\begin{center}
$\gamma = (87 ^{+11} _{-12})^{\circ}$
\end{center}
where the uncertainty corresponds to the 68 \% confidence interval.
This is the most precise determination of $\gamma$ from a single analysis. At present the
result is statistically limited but the analysis presented was using only 2015--2016 data, 
so only 2 out of the 5.9 fb$^{-1}$ available in Run2.

\section{$B^+ \rightarrow D K^{\star +}$ (2, 4 body D decays) analysis}

In the decay mode {$B^+ \rightarrow D K^{\star +}$, where the neutral $D$ meson decays
to $h^+ h^-$ and $h^+ \pi- \pi+ \pi-$, $h = K, \pi$ and the $K^{\star +}$ meson to
$K^0_{\text{S}} \pi^+$, the sensitivity to $\gamma$ is obtained from the interference observed 
by reconstructing the $D$ meson in final states accessible to both $D^0$ and 
$\overline{D}^0$ \cite{ref:input2}. Up to 12 CP observables can be measured. For 
illustration purposes, one of them is defined as:
\begin{center}
$A_{KK} \equiv \frac{\Gamma(B^- \rightarrow D(K^+K^-) K^{\star -})-\Gamma(B^+ \rightarrow D(K^+K^-) K^{\star +})}
{\Gamma(B^- \rightarrow D(K^+K^-) K^{\star -})+\Gamma(B^+ \rightarrow D(K^+K^-) K^{\star +})}$
\end{center}
which represents the CP asymmetry for the $D \rightarrow K^+K^-$ decay mode.
$A_{\pi\pi}$ is defined similarly but swapping $K^+K^-$ with $\pi^+ \pi^-$.
As direct CP violation in $D$ decay is small $A_{KK} = A_{\pi\pi} \equiv A_{CP+}$ where
\begin{center}
$A_{CP+} = \frac{2 k r_B \sin\delta_B \sin\gamma}{1+r_B^2+ 2 k r_B \cos\delta_B \cos\gamma}$
\end{center}
with $r_B$ and $\delta_B$ defined as in the previous section and $k$ a dilution factor 
for non $K^{\star}(892)^- \rightarrow K^0_{\text{S}} \pi^-$  contributions. This shows the connection, for
one of the 12 CP observables, with the physical parameter of interest, $\gamma$.
The observed invariant mass distribution for $B^- \rightarrow D(K^+K^-) K^{\star -}$ is shown in
the left plot of Fig. \ref{fig:input2_1} with the charge conjugated process on the right. 
\begin{figure}[h]
\centering
\includegraphics[width=80mm]{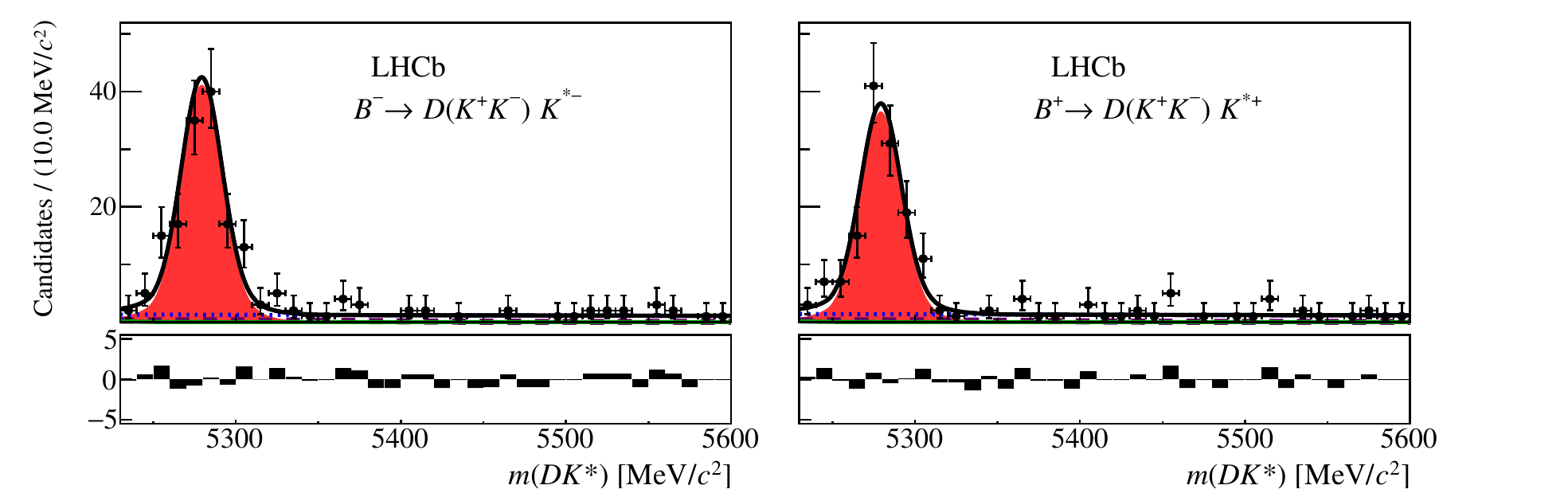}
\caption{Observed invariant mass distribution for (right) $B^- \rightarrow D(K^+K^-) K^{\star -}$ and
(left) $B^+ \rightarrow D(K^+K^-) K^{\star +}$. Reproduced from Ref. \cite{ref:input2}.}
\label{fig:input2_1}
\end{figure}
Overall, 7 different $D$ decay modes are considered with the observed yields in the $B^+$
and $B^-$ cases reported in Tab. \ref{tab:input2_2}. The analysis is using the 2011 to 2016 
LHCb data set.
\begin{table}[h]
\begin{center}
\caption{$B^+$ and $B^-$ measured yields for each of the 7 considered $D$ decay modes.                              
Reproduced from Ref. \cite{ref:input2}.}
\begin{tabular}{|l|c|c|}
\hline Decay mode & $B^-$ yield & $B^+$ yield \\
\hline $B^{\pm} \rightarrow D(K^{\pm} \pi^{\mp}) K^{\star \pm}$             & 996 $\pm$ 34 & 1035 $\pm$ 35 \\
\hline $B^{\pm} \rightarrow D(K^+ K^-) K^{\star \pm}$                       & 134 $\pm$ 14 & 121  $\pm$ 13 \\
\hline $B^{\pm} \rightarrow D(\pi^+ \pi^-) K^{\star \pm}$                   & 45  $\pm$ 10 & 33   $\pm$ 9  \\
\hline $B^{\pm} \rightarrow D(K^{\mp} \pi^{\pm}) K^{\star \pm}$             & 1.6 $\pm$ 1.9& 19   $\pm$ 7  \\
\hline $B^{\pm} \rightarrow D(K^{\pm} \pi^{\mp} \pi^+ \pi^-) K^{\star \pm}$ & 556 $\pm$ 26 & 588  $\pm$ 27 \\
\hline $B^{\pm} \rightarrow D(\pi^+ \pi^- \pi^+ \pi^-) K^{\star \pm}$       & 59  $\pm$ 10 & 56   $\pm$ 10 \\
\hline $B^{\pm} \rightarrow D(K^{\mp} \pi^{\pm} \pi^- \pi^+) K^{\star \pm}$ & 3   $\pm$ 5  & 10   $\pm$ 6  \\
\hline
\end{tabular}
\label{tab:input2_2}
\end{center}
\end{table}
These numbers allow the extraction of the 12 CP observables. For illustration purposes:
\begin{center}
$A_{KK} = -0.004 \pm 0.023 \pm 0.008$ \\
$A_{\pi \pi} = 0.15 \pm 0.13 \pm 0.02$
\end{center}
where the first uncertainty is statistical and the second systematic. As anticipated,
within the present uncertainties, $A_{KK}$ and $A_{\pi \pi}$ are indeed equal.
Please see Ref. \cite{ref:input2} for a full summary of the results.
The physical interpretation in term of $r_B$ and $\gamma$, using the full set of CP observables,
is given in Fig. \ref{fig:input2_3}.
\begin{figure}[h]
\centering
\includegraphics[width=80mm]{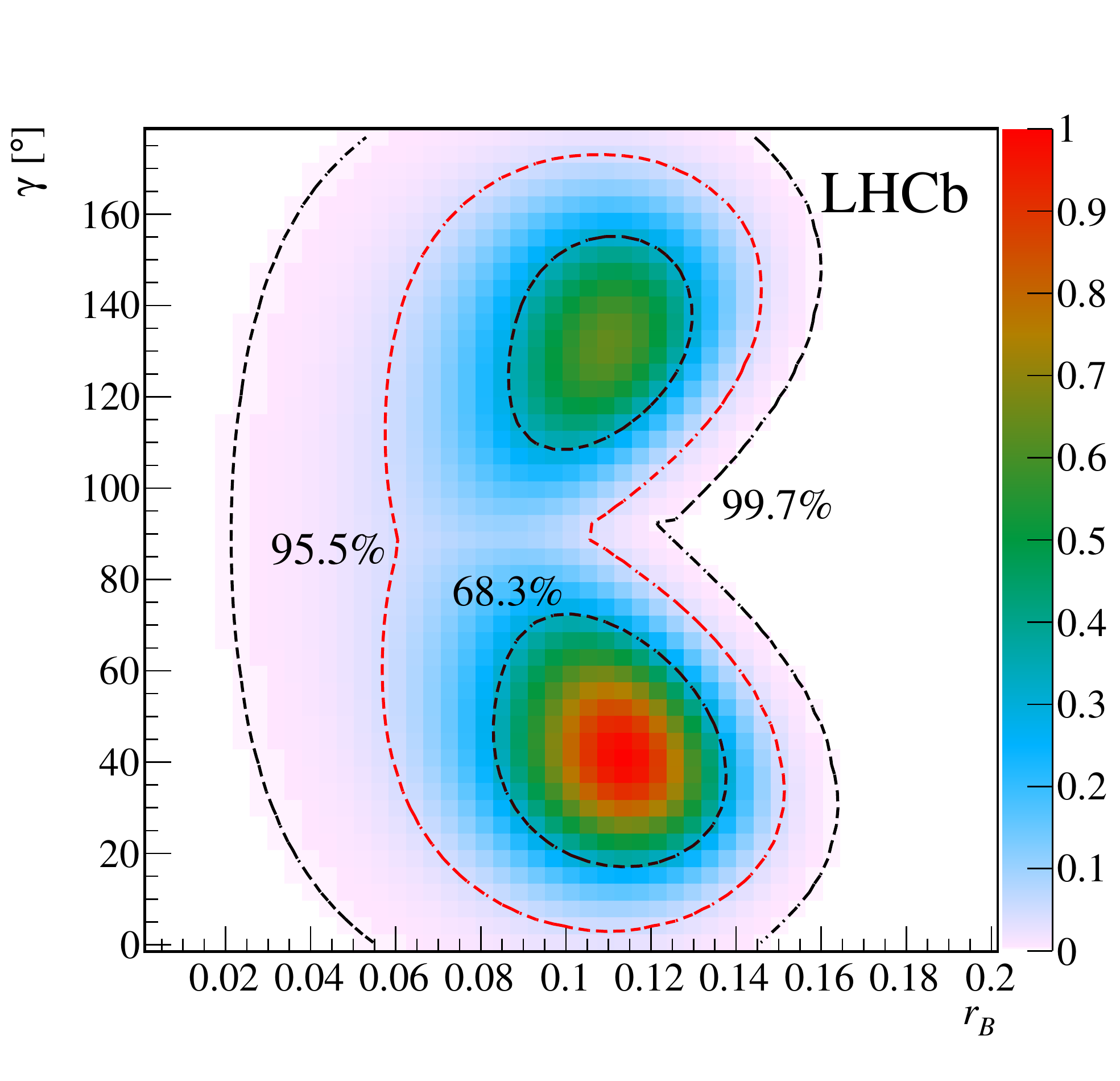}
\caption{$r_B$ vs $\gamma$ allowed region using the values of the 12 CP observables from
$B^{\pm} \rightarrow D K^{\star \pm}$. Reproduced from Ref. \cite{ref:input2}.}
\label{fig:input2_3}
\end{figure}
Alone this particular decay mode has a limited sensitivity but results are
consistent with $\gamma \sim 70^{\circ}$ and $r_B \sim 0.1$. Measurements are statistically
limited but statistic will increase from 5.2 to 9.1 fb$^{-1}$ once the 2017 and the 2018 data 
samples are included.

\section{$B^0_s \rightarrow D_s K$ analysis}

In the decay mode $B^0_s \rightarrow D_s^{\mp} K^{\pm}$ the sensitivity to $\gamma$ is 
obtained from the interference of decay amplitudes with and without mixing \cite{ref:input3}.
This is a time dependent analysis requiring flavour tagging to determine the flavour of the
reconstructed neutral $B$ meson at production time. The CP parameters related
to $r_B$, $\delta_B$ and $(\gamma - 2 \beta_s)$, 
where $\beta_s \equiv \arg( - V^{\phantom{\star}}_{ts}V^{\star}_{tb}/V^{\phantom{\star}}_{cs}V^{\star}_{cb})$ , 
are obtained fitting the observed decay time distribution.
The fit results to the $D_s^{\mp} K^{\pm}$ invariant mass distribution are illustrated in
Fig. \ref{fig:input3_1}. Using the 2011--2012 data sample a signal yield of 5955 $\pm$ 90
is obtained.
\begin{figure}[t]
\centering
\includegraphics[width=80mm]{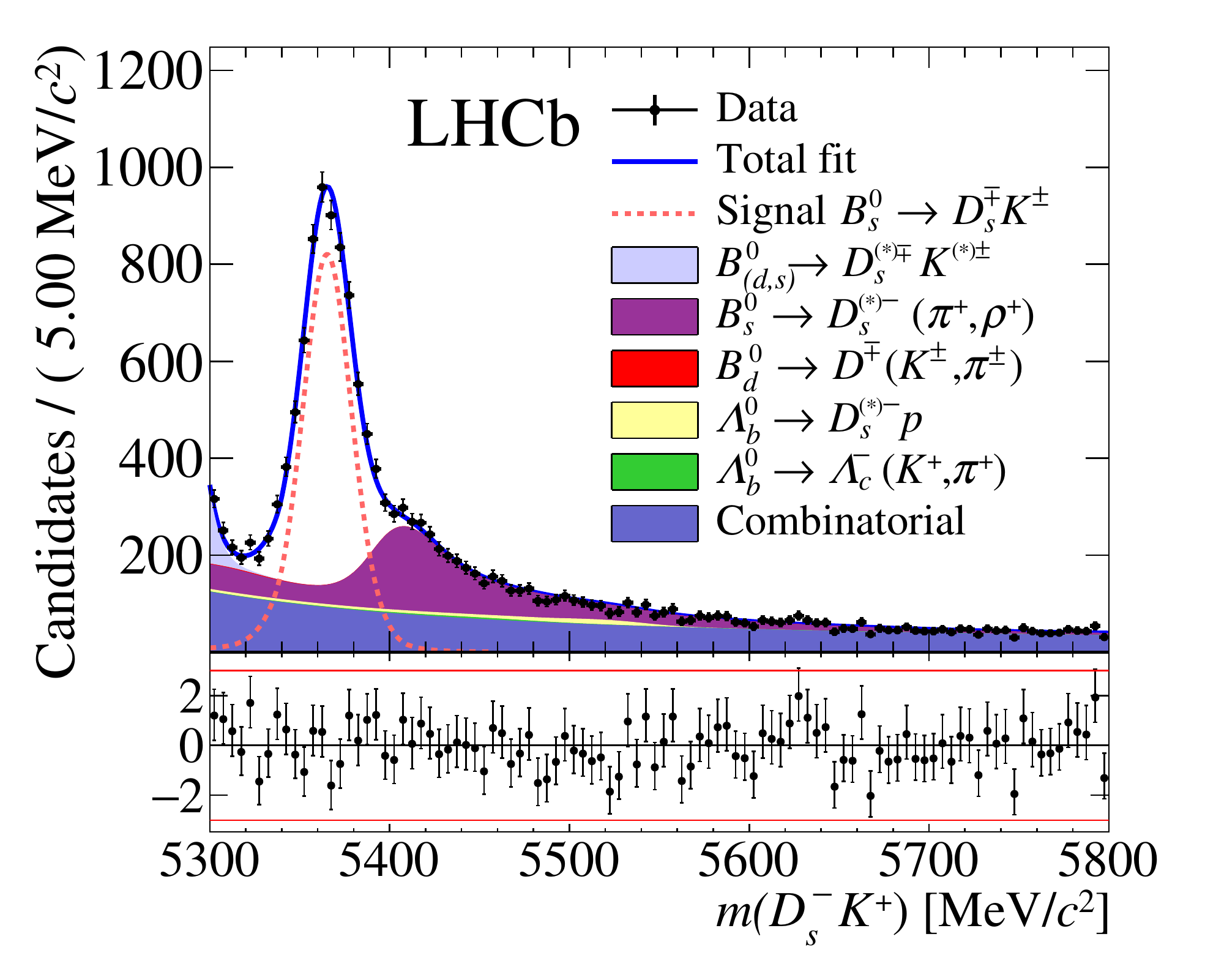}
\caption{$B^0_s \rightarrow D_s^{\mp} K^{\pm}$ invariant mass distribution.
Reproduced from Ref. \cite{ref:input3}.}
\label{fig:input3_1}
\end{figure}
The observed decay time distribution is shown in Fig. \ref{fig:input3_1}. The red
dashed line in the same figure corresponds to the decay time acceptance as obtained
from $B^0_s \rightarrow D_s^- \pi^+$ data after a small correction
obtained from the $D_s^{\mp} K^{\pm}$ to $D_s^- \pi^+$ time acceptances ratio as
obtained from Monte Carlo samples. The CPV parameters fit result is given by the blue 
continous line.
\begin{figure}[b]
\centering
\includegraphics[width=80mm]{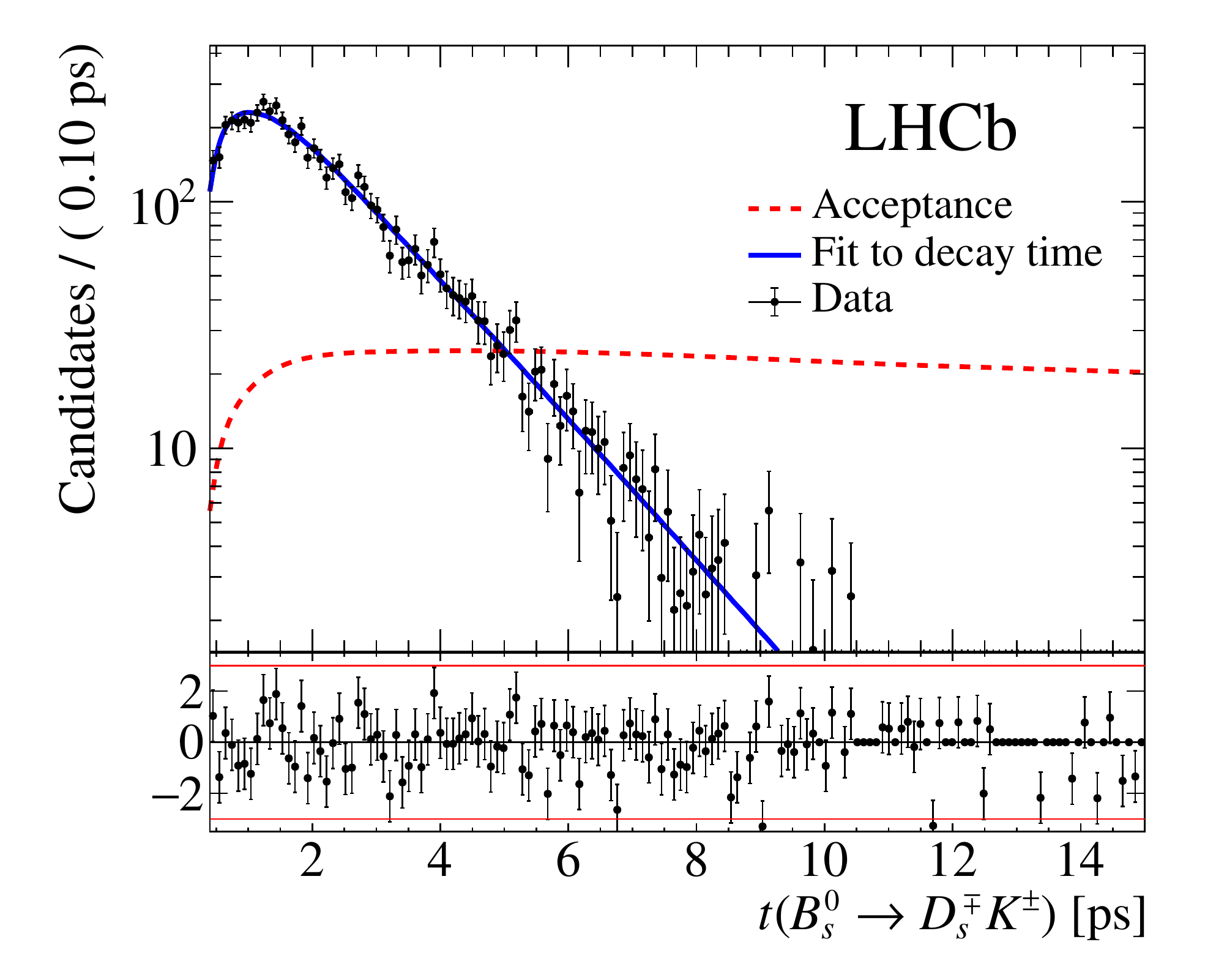}
\caption{Fit to the $B^0_s \rightarrow D_s^{\pm} K^{\mp}$ observed decay time
distribution. Reproduced from Ref. \cite{ref:input3}.}
\label{fig:input3_2}
\end{figure}
As the effect of CPV is difficult to appreciate in Fig. \ref{fig:input3_2}, the
folded time asymmetries for $D_s^+ K^-$ and $D_s^- K^+$ are shown in the left and
right plots of Fig. \ref{fig:input3_3}, respectively. The effect of CPV can
then be seen as a phase difference between the two asymmetries different from $\pi$ at 
$t = 0$ ps. The final result for $\gamma$ from this analysis being:
\begin{center}
$\gamma = (128 ^{+17}_{-22})^{\circ}.$
\end{center}
This is the most precise determination of $\gamma$ from a $B_s$ meson decay. The
result is obtained using so far only the 3 fb$^{-1}$ collected in 2011--2012 and
will be extended to the 5.9 fb$^{-1}$ collected during Run2 allowing to
improve significantly the statistical accuracy on $\gamma$.
\begin{figure}[h]
\includegraphics[height=30mm]{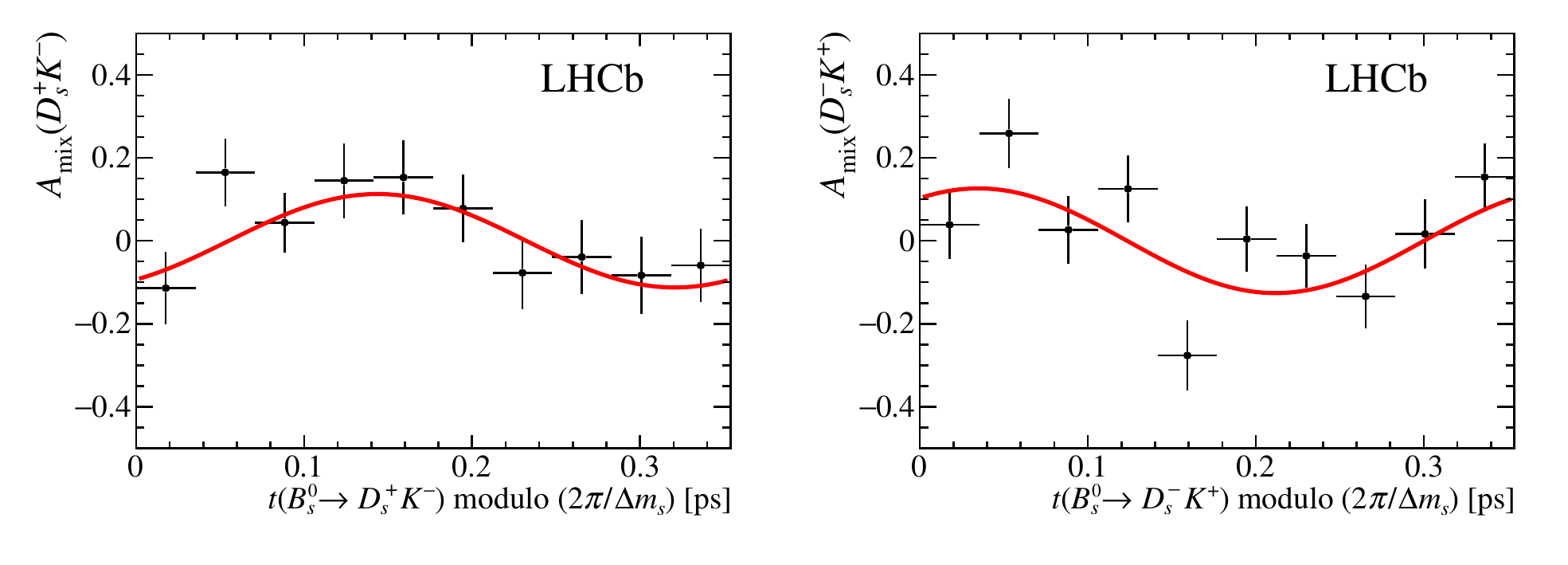}
\caption{Folded time asymmetry for (left) $D_s^+ K^-$ and (right) $D_s^- K^+$.
Reproduced from Ref. \cite{ref:input3}.}
\label{fig:input3_3}
\end{figure}

\section{$\gamma$ combination results}

As the most precise determination of $\gamma$ from a single measurement
presently has a statistical uncertainty around 10$^{\circ}$, which is large
with respect to what desirable, it is mandatory
to combine the measurements obtained from all the accessible decay 
modes \cite{ref:LHCb-CONF-2018-002}.
The present LHCb $\gamma$ combination uses as input 98 observables to constrain
40 free parameters. The main results of the fit are $\gamma$, treated as a common
free parameter, and the $r_B$ and $\delta_B$ values of each considered
decay mode.
In order to extract $\gamma$ from the measurements presented in the previous
sections some "auxiliary" inputs are also needed. One example being the value of
$\beta_s$ for the $B^0_s \rightarrow D_s K$ measurement.
Whenever possible these auxiliary inputs are taken from data, whenever possible from
LHCb data. These are Gauss--constrained in the combination. Treating them as free
parameters roughly doubles the uncertainty on $\gamma$.
The combination result for $\gamma$ is:
\begin{center}
$\gamma = (74.0^{+5.0}_{-5.8})^{\circ}$
\end{center}
where the accuracy increases by a factor of about 2 with respect to the most accurate
single measurement. The 1-CL curve for the $\gamma$ parameter is shown in 
Fig. \ref{fig:input4_1} with central value (solid vertical line) and 1 $\sigma$
uncertainties (dashed vertical lines) labelled. The 1 and 2 $\sigma$ levels are 
indicated by the horizontal dotted lines.
\begin{figure}[h]
\centering
\includegraphics[width=80mm]{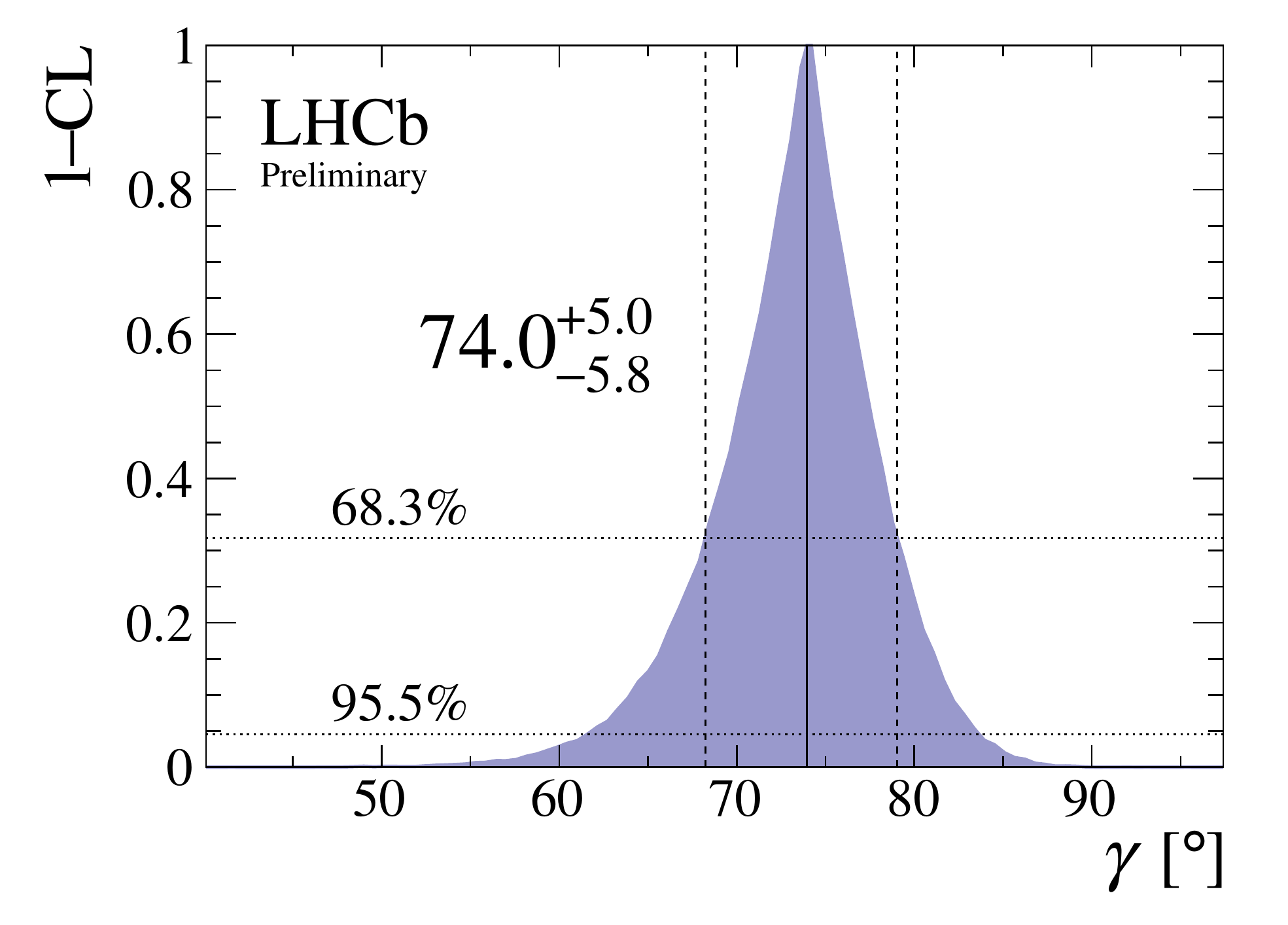}
\caption{1-CL curve for the $\gamma$ combination. Reproduced from 
Ref. \cite{ref:LHCb-CONF-2018-002}.}
\label{fig:input4_1}
\end{figure}
$B^+$, $B^0$ and the single $B^0_s$ meson results are currently used in
the $\gamma$ combination, the corresponding 1-CL curves are shown in violet,
yellow and orange in Fig. \ref{fig:input4_2}, respectively. The green curve shows
the combination is as in Fig. \ref{fig:input4_1}. Such a "differential" analysis
allows to probe the stability and the strength of the result.
$B^+$ are clearly driving the final result. $B^0$ and $B^0_s$ are subdominant
and having almost the same weight. This happens because the $r_B$ value of the
single $B^0_s$ measurement is 0.301, the largest measured so far.
Future measurements using $B_c^{\pm} \rightarrow D_s^{\pm} D$, the analogue of
$B^0_s \rightarrow D_s^{\pm} K^{\mp}$ replacing the $s$ quark with a $b$ quark, 
penalized by a small production yield, are expected to have $r_B \sim O(1)$
\cite{ref:GMK}.
\begin{figure}[h]
\centering
\includegraphics[width=80mm]{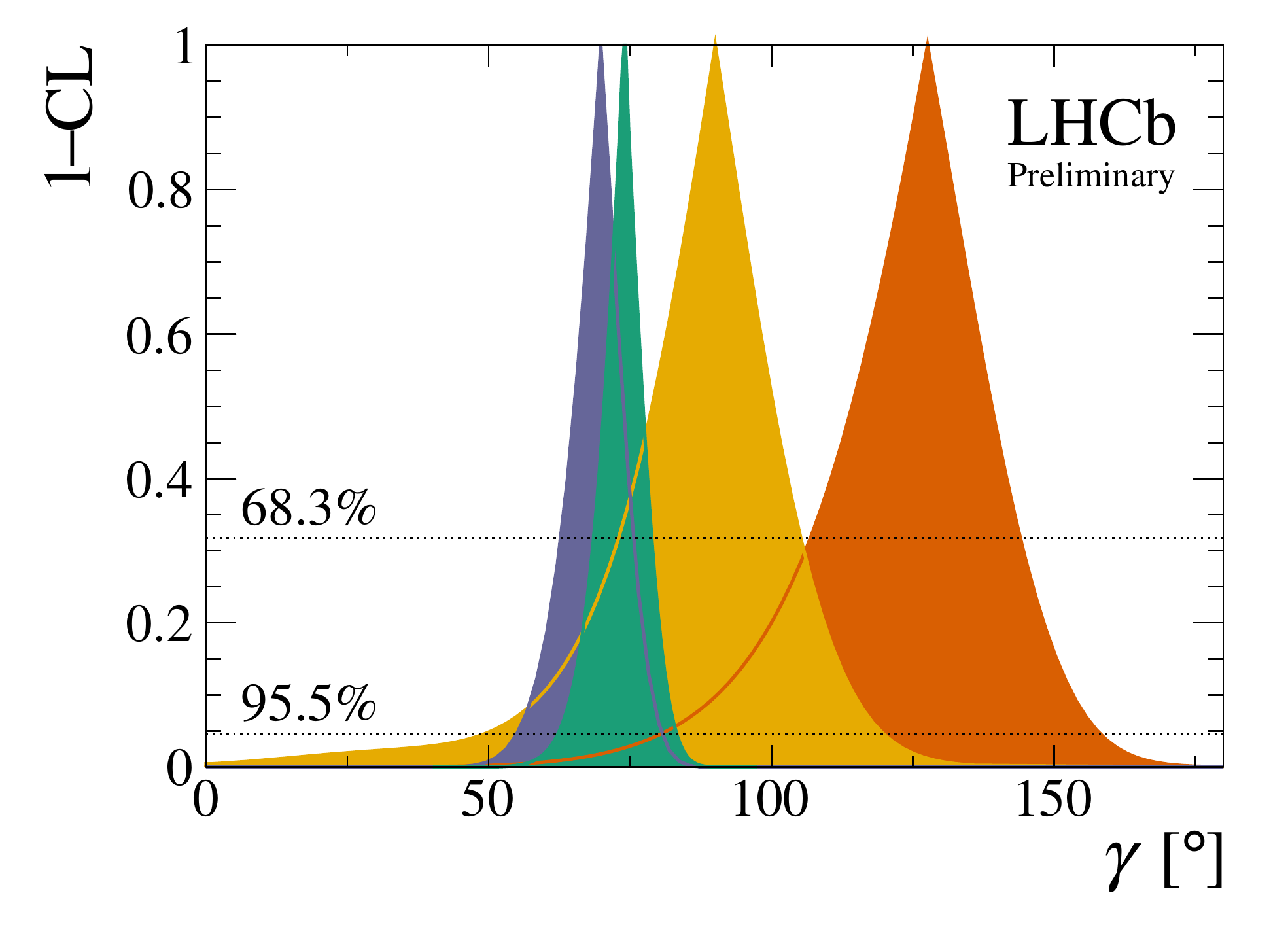}
\caption{1-CL plots for combinations split by the initial B meson flavour.
Reproduced from Ref. \cite{ref:LHCb-CONF-2018-002}.}
\label{fig:input4_2}
\end{figure}

\section{Summary and future prospects}

The result of the 2018 LHCb $\gamma$ combination is:
\begin{center}
$\gamma = (74.0^{+5.0}_{-5.8})$
\end{center}
based on 16 input measurements that cross check each other and allow to evaluate 
the stability of the combined result. This value is compared in Fig. \ref{fig:summary}
with the CKMfitter \cite{ref:CKMfitter} and UTfit \cite{ref:UTfit}  global fit results, 
as of Summer 2018.
Clearly the present LHCb uncertainty on $\gamma$ does not yet allow to draw any stringent
conclusion from the comparison between tree--level and global fits determinations.
On a short term time scale LHCb will extend all input measurements to the full Run1 plus
Run2 data set. In addition new measurements are about to come at the time these proceedings 
are being written. Longer term, starting from about 2021, new data will be
available thanks to the high luminosity LHC upgrade. With possibly additional inputs
to the combination.
Current projections indicate that with a luminosity of 23 fb$^{-1}$, available by about 2024, 
an accuracy of 1.5$^{\circ}$ should be reachable. As shown in Fig. \ref{fig:summary}, at that
time the LHCb uncertainty will be similar to the present global fits uncertainty (in Fig. 
\ref{fig:summary} the central value of the expected 23 fb$^{-1}$ result has been arbitrarily 
kept to its current value). If the accuracy of the external inputs will not limit the LHCb
measurements and if a luminosity of 300 fb$^{-1}$ could be reached by the end of LHC operations
the uncertainty on $\gamma$ should shrink to 0.35$^{\circ}$. 
\begin{figure}[t]
\centering
\includegraphics[width=80mm]{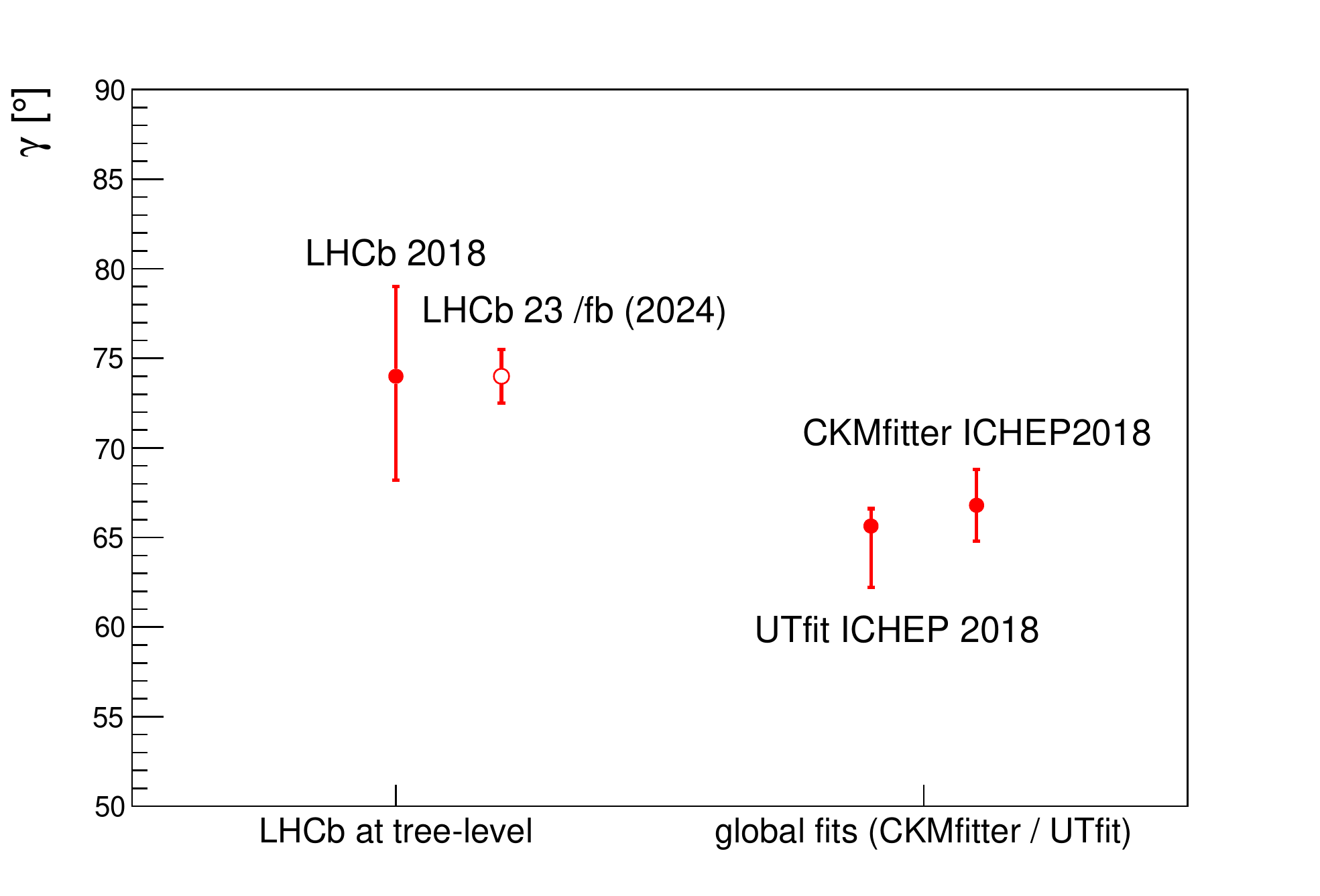}
\caption{Comparison between the present LHCb $\gamma$ result from tree--level decays and the
global fit determinations from CKMfitter and UTfit. A projection of the LHCb $\gamma$ result
based on the data sample that will be collected by 2024, after the high luminosity LHC upgrade,
is also shown.}
\label{fig:summary}
\end{figure}
At present LHCb has improved the accuracy of the $\gamma$ measurement obtained by BaBar
or Belle by a factor of about 3.
In the forthcoming months Belle II, that started full physics operation in 2019, will start
to push towards a reduction of the uncertainty on $\gamma$, having about the same expected 
sensitivity as LHCb.


\bigskip 

\end{document}